# Giant spin-charge conversion in ultrathin films of the MnPtSb half-Heusler compound

E. Longo[1,*], A. Markou[2], C. Felser[2], M. Belli[1], A. Serafini[3], P. Targa[3], D. Codegoni[3], M. Fanciulli[4], R. Mantovan[1,*]

1. *CNR-IMM, Unit of Agrate Brianza (MB), Via C. Olivetti 2, 20864, Agrate Brianza (MB), Italy*
2. *Max Planck Institute for Chemical Physics of Solids, Nöthnitzer Straße 40, Dresden, 01187, Germany*
3. *STMicroelectronics, Via C. Olivetti 2, 20864 Agrate Brianza, Italy*
4. *Università degli studi di Milano-Bicocca, Dip. di Scienze dei Materiali, Via R. Cozzi 55,20126, Milano, Italy*

*Corresponding authors: emanuele.longo@mdm.imm.cnr.it \*, roberto.mantovan@cnr.it \*\**

**Abstract**

Half-metallic half-Heusler compounds with strong spin-orbit-coupling and broken inversion symmetry in their crystal structure are promising materials for generating and absorbing spin-currents, thus enabling the electric manipulation of magnetization in energy-efficient spintronic devices. In this work, we report the spin-to-charge conversion in sputtered ultrathin films of the half-Heusler compound MnPtSb with thickness ($t$) in the range from 1 to 6 nm. A combination of X-ray and transmission electron microscopy measurements evidence the epitaxial nature of these ultrathin non-centrosymmetric MnPtSb films, with a clear (111)-orientation obtained on top of (0001) single-crystal sapphire substrates. The study of the $t$-dependent magnetization dynamics of the MnPtSb(t)/Co(5nm)/Au(5nm) heterostructure revealed that the MnPtSb compound can be used as an efficient spin current generator, even at film thicknesses as low as 1 nm. By making use of spin pumping FMR, we measure a remarkable $t$-dependent spin-charge conversion in the MnPtSb layers, which clearly demonstrate the interfacial origin of the conversion. When interpreted as arising from the inverse Edelstein effect (IEE), the spin-charge conversion efficiency extracted at room temperature for the thinnest MnPtSb layer reaches $\lambda_{IEE} \sim 3\ nm$, representing an extremely high spin-charge conversion efficiency at room temperature. The still never explored ultrathin regime of the MnPtSb films studied in this work and the discover of their outstanding functionality are two ingredients which demonstrate the potentiality of such materials for future applications in spintronics.

**Introduction**

The need to control the energy consumption and the development of highly efficient electronic devices, in terms of *in operando* speed and occupation on a single chip, has recently become of central interest for the industrial players of the semiconductor market.[1–5] In the last decades, spintronic devices emerged as one of the most promising options to shape the future of more-efficient and lower-power consuming nanoelectronics.[6–8] Within spintronics, new physical paradigms, novel materials, and systems allowing to efficiently manipulate spin-currents and their interconversion with conventional charge-currents, are continuously under the radar. In particular, to optimize spin-charge conversion (SCC) processes is of particular importance, being a mechanism at the core of the functionality of several spintronic devices.[5–9] The first studies concerning SCC-based electronic devices were focused on the coupling between ferromagnetic (FM) materials and heavy metals (i.e. Pt, Pd, Ta). The interest is gradually shifting to alternative materials such as topological insulators,[9,10] Weyl semimetals,[11–13] transition metal dichalcogenides,[14] and 2D-materials[14,15] with the general aim of enhancing the SCC efficiency. Very recently, Heusler compounds (HC) also entered the game of spin-charge interconversion.[16–18] HC is a class of intermetallic ternary compounds discovered in 1903 by Fritz Heusler, which comprehends HC and half-HC, with a 2:1:1 or 1:1:1 stoichiometry, respectively. The renewed interest in HC and half-HC arose from their versatility in terms of both magneto-optical properties and chemical composition (more than 1500 combinations exist), which makes possible the formation of semiconductors, metals, topological materials, or superconductors.[18] A sub class of HC and half-HC is characterized by a non-trivial band structure which makes them half-metallic compounds, having an ideally 100% spin-polarized transport.[19,20] In principle, in the case of ideal 100% spin polarization at the Fermi level, these materials should not allow spin accumulation, therefore preventing their use as efficient sources of spin currents and/or spin-charge converters.[17,18,20] However, theoretical calculations have predicted that SCC can occur in NiMnSb, within a magnon-assisted mechanism, which is ultimately originated by a deviation from the ideal half-metallicity that can occur in real materials.[17]

Moreover, HC and half-HC possess a non-centrosymmetric crystal structure which promotes the formation of topological states, which makes them viewable as a kind of topological insulators or Weyl semimetals either, being a very promising aspects to consider for enhancing the SCC in many systems.[18,21,22] One of the first attempts in making used of HC for spin-charge interconversion applications has been reported by Ciccarelli *et al.*[16] making use of a epitaxial NiMnSb thin film which was employed as ferromagnetic layer to conduct spin orbit torque experiments at room temperature. In that work, by a combined experimental and theoretical work, the authors settled the origin of the measured spin torque signals, disentangling the predominant contribution of the Dresselhaus field from the less influent Rashba field. Differently, Z. Wen *et al.*[17] observed surface and bulk contributions in the SCC extracted by polycrystalline NiMnSb-based heterostructures by means of spin pumping measurements. Interestingly, the authors performed temperature dependent measurements, which allowed to attribute the surface SCC as due to the presence of the Inverse Edelstein Effect (IEE) in the NiMnSb films.

Here, we present the new ultrathin half-HC MnPtSb system, which we demonstrate to generate an unprecedented SCC efficiency at room temperature. Cubic MnPtSb thin films are deposited on top of a [0001] oriented sapphire substrate at variable thickness t = 1, 1.5, 2, 3, 4, 5, 6 nm. The whole set of samples is capped with 5 nm Co and 5 nm Au layers, to produce the final $Al_2O_3$/MnPtSb(t)/Co(5)/Au(5) heterostructures. All samples were investigated by X-ray diffraction (XRD) and transmission electron microscopy (TEM) measurements to fully reveal their chemical-structural properties. The experimental investigation of the samples' magnetization dynamics was conducted through broadband ferromagnetic resonance (BFMR) spectroscopy. In order to probe the SCC efficiency occurring in the developed MnPtSb(t)-based heterostructures, spin pumping ferromagnetic resonance (SP-FMR) experiments were conducted. Based on BFMR and SP-FMR results, the origin of the SCC in the MnPtSb-based heterostructures is found to be intimately guided by the MnPtSb/Co interface, with a clear increase of SCC at lower thicknesses. Within the IEE type of conversion, the extremely high IEE length of $\lambda_{IEE} \sim 3\ nm$ is measured at room temperature for the 1 nm thick MnPtSb system. It is important to stress that no interlayer was used at the MnPtSb/Co interface, showing the robustness of such interface against the presence of structural disorder and chemical intermixing. The summary of all the structures is reported in Table 1 together with some of the parameters of interest.

**MnPtSb structural properties**

    a. **X-ray diffraction (XRD)**

Figure 1(a) displays the non-centrosymmetric unitary cell of the half-HC compound MnPtSb, which belongs to the F-43m space group.[23] To determine the crystalline orientation of the sputtered MnPtSb films, we carried out XRD measurements in the Bragg-Brentano geometry, allowing to identify the orientation of the films along the out-of-plane (OOP) direction and evaluate their mosaicity.[24,25] (see Methods and Supporting information).

Figure 1(b) depicts the $\omega - 2\vartheta$ maps of the MnPtSb(6) heterostructures, where an intense signal is observed at $2\theta \sim 25°$, which we attribute to the OOP MnPtSb (111) crystalline planes.[23] Here, the limited broadening along the $\omega$ axis of the red signal indicates a high degree of crystallinity of the deposited MnPtSb film, suggesting a nearly epitaxial fashion. The high crystalline quality of the MnPtSb(6) sample is also confirmed by the presence of the Au (111) reflection around $2\theta \sim 38°$ on the Bragg-Brentano line (i.e. bisector of the $\omega - 2\vartheta$ maps, which indicates that the Au capping layer grows epitaxially along the (111) OOP orientation, thus following the MnPtSb crystalline structure. The Co reflections are not visible in the $\omega - 2\vartheta$ maps due to the very high intensity of the MnPtSb signal. In addition, the high fluorescence arising from Au, increases the signal background, hiding the reflections with lower intensity.[25]

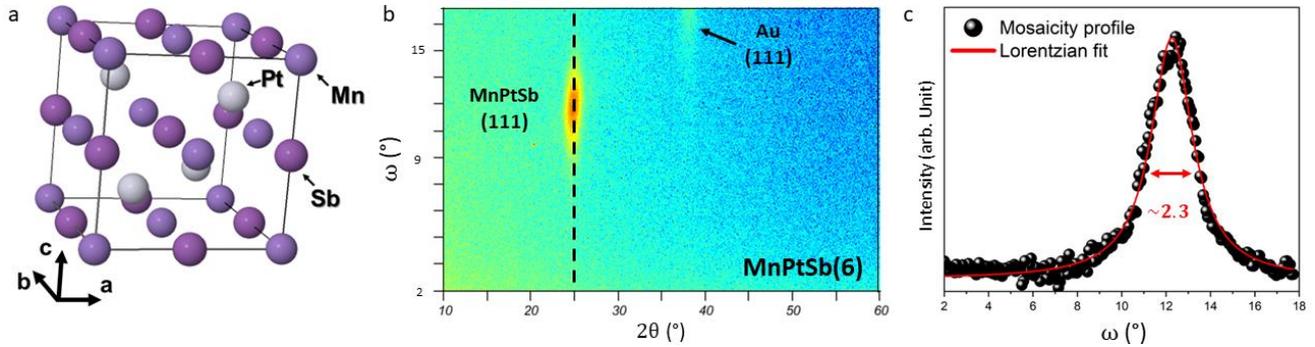

**Figure 1:** (a) Representation of the crystalline structure of the half-Heusler MnPtSb compound (ICSD Code: 628387). (b) XRD measurements performed in Bragg-Brentano geometry to determine the OOP orientation of the MnPtSb substrate for the MnPtSb(6) sample. (c) Mosaicity profile extracted along the black dashed line in panel (b).

Figure 1(c) represents the intensity profile acquired along the black dashed line drawn in Fig. 1(b), which has a typical Lorentzian shape. The full width at half maximum of such a peak represents the mosaicity degree of the MnPtSb(6) film, a quantity that gives a measure of the deviation of the [111] direction of the MnPtSb film from the ideal OOP orientation. The measured 2.3° value is very low if compared to that typically observed in ultra-thin sputtered films, thus confirming an almost epitaxial character of our MnPtSb layers.[26,27]

## b. Transmission electron microscopy (TEM) and Energy Dispersive X-ray (EDX) analyses

To gain more insights about the crystallinity and elemental distribution within the MnPtSb films and whole heterostructure, we conducted TEM and energy dispersive X-ray analysis (EDX) measurements for selected MnPtSb layers, with t = 6, 2, and 1 nm. In addition, we grew $Al_2O_3$/MnPtSb(6)/$SiO_x$(3) and $Al_2O_3$/MnPtSb(2)/$SiO_x$(3) films on sapphire, in order to judge the MnPtSb films quality on the nanoscale (see Fig. S1 in the Supplementary Information). The results are depicted in Fig.2, where panels (a), (b) and (c) show the atomic structure of the MnPtSb for t = 6, 2, 1 nm respectively, and the yellow squares indicate the regions in which the Fast Fourier transformation (FFT) is performed (see insets in corresponding panels). As it is clear from TEM images and corresponding FFT patterns, the epitaxial MnPtSb are discontinuous. Among all the samples, the MnPtSb(6) exhibits high crystallinity in agreement with XRD (Fig.1), while MnPtSb(2) and MnPtSb(1) show a more polycrystalline nature. We also conducted EDX on MnPtSb(6), MnPtSb(2) and MnPtSb(1), and the results are seen in Fig. 2 (d), (e) and (f), respectively. Here, the different colors indicate the signals corresponding to all the probed chemical elements as indicated in the insets of panels (d), (e) and (f). Due to the island-growth mode of the MnPtSb films in such an ultrathin regime, the cutting procedure to prepare lamellas for the TEM analysis takes place over an area where different local stacking of the elements appears. This makes the TEM analysis very challenging, especially for estimating the potential interdiffusion between atoms from adjacent layers. On the other hand, it is quite straightforward to get insight about the elemental distribution inside the MnPtSb layers of varying thickness. Figures 2 (g), (h) and (i) display the elemental distribution profile as ideally taken along the dashed lines drawn in Fig. 2 (d), (e), (f), for the Mn (red squares), Pt (yellow triangles), and Sb (green circles) atoms. Actually, the profiles in Fig. 2 (g), (h), (i), are all averages over four separate vertical profiles of the TEM lamella, in order to better represent the overall films' character. By following the different EDX profiles, it can be noticed that the Mn signal is slightly shifted towards the $Al_2O_3$ substrate in all heterostructures, likely promoting the formation of a PtSb-rich region at a close vicinity of the Co layer. Due to the different thicknesses of the MnPtSb layer in samples MnPtSb(6), MnPtSb(2) and MnPtSb(1), the stoichiometric modification of the MnPtSb compound appears more relevant at the lower thicknesses of t = 1 and 2 nm.

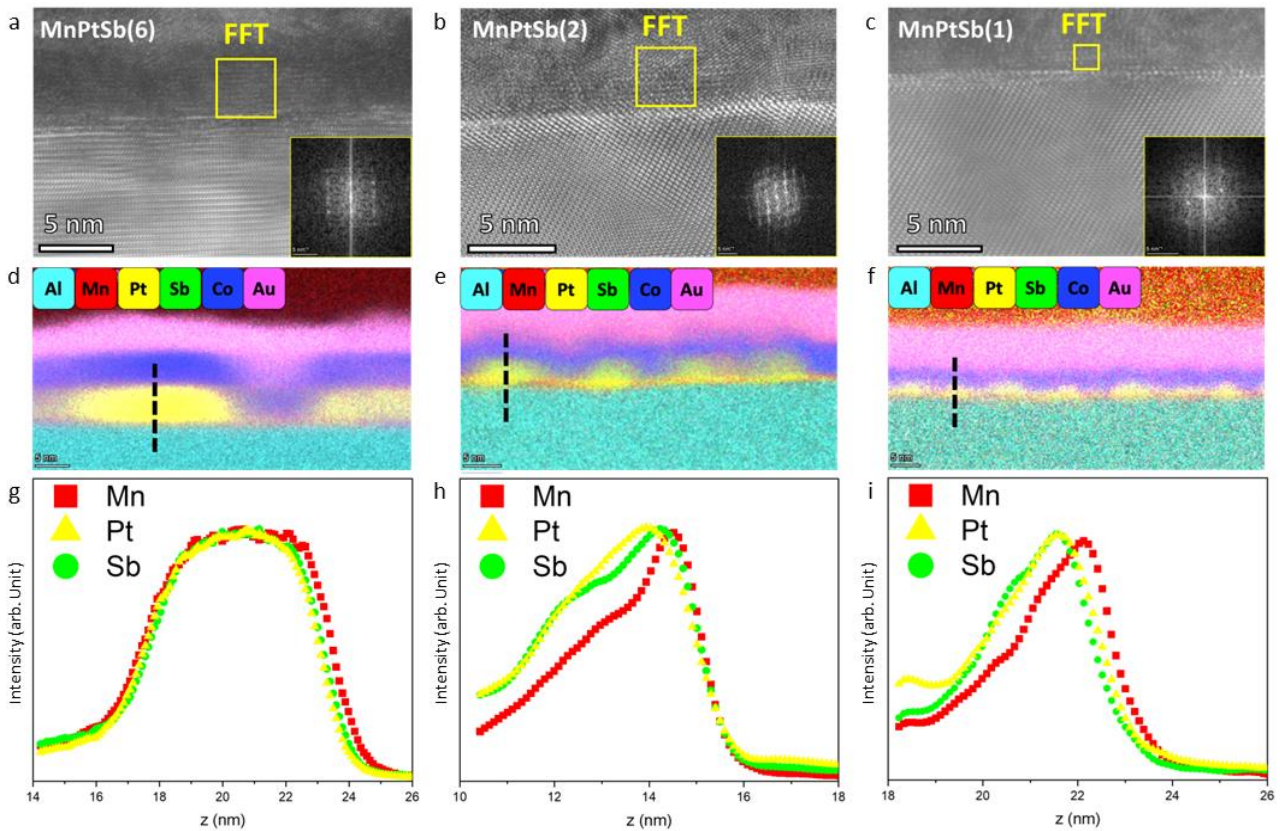

**Figure 2:** In panels (a), (b) and (c) the crystalline structure as extracted from the TEM measurement for samples MnPtSb(6), MnPtSb(2) and MnPtSb(1) are reported. In the inset of these panels the FFT images of the regions highlighted by the yellow square are shown for each sample. In panels (d), (e) and (f) the EDX images are reported for the same samples. Here, the black dashed lines indicate the position in which the elemental profiles showed in panels (g), (h), (i) are acquired.

**Broadband Ferromagnetic Resonance (BFMR) and spin pumping characterization**

The magnetization dynamics in MnPtSb/Co/Au heterostructures were investigated using BFMR, while varying the angle between the applied external magnetic field and a reference direction in the sample plane (IP, in-plane configuration). A MnPtSb-free $Al_2O_3$/Co/Au sample (named "REF" in the following) was grown to serve as a reference, a necessary step to evaluate the influence of the MnPtSb layer in the generation of spin currents. The time evolution of the magnetization vector $(\vec{M})$ in a magnetic material under the effect of an effective external magnetic field $(\vec{H}_{ext})$ is described by the Landau-Lifshitz-Gilbert equation[28] (Eq.1)

$$\frac{d\vec{M}}{dt} = -\mu_0 \gamma \vec{M} \times \vec{H}_{ext} + \frac{\alpha}{M_s}\left(\vec{M} \times \frac{d\vec{M}}{dt}\right) \qquad [1]$$

Here, $\mu_0$ represents the magnetic permittivity, $\gamma = \frac{ge}{2m_e} = \frac{g\mu_B}{\hbar}$ the gyromagnetic ratio (where $g$ is the g-factor, $e$ the electron charge, $m_e$ the effective mass for the free electron, $\mu_B$ the Bohr magneton and $\hbar$ the reduced Plank constant), and $\alpha$ the damping constant. In the specific case in which $\vec{H}_{ext}$ lies IP, it is possible to link the resonant frequency $f_{res}$ of the magnetization with the resonant magnet field $(H_{res})$ through the Kittel equation[29] that holds in the case of IP configuration (Eq.2), as depicted in Eq. 2.[30]

$$f_{res} = \frac{\gamma}{2\pi}\sqrt{(H_{res} + H_{k1}(\varphi))(H_{res} + H_{k2}(\varphi) + 4\pi M_{eff})} \qquad [2]$$

where $\varphi$ is the IP angle between $\vec{H}_{ext}$ and the easy axis of the magnetization vector (see inset of Fig. 3 (a)). $H_{k1}$ and $H_{k2}$ represent the IP anisotropy magnetic fields, which embed all the possible contributions arising from different crystal symmetries. We obtained $H_{res}(\varphi)$ curves at a fixed frequency to investigate the presence of the IP anisotropy. The results revealed a significant uniaxial anisotropy in the Co layers on top of all MnPtSb films, where the easy and hard axes directions are perpendicular to each other. This is demonstrated by the presence of a 180° periodic oscillation in the collected $H_{res}(\varphi)$ curves. For more details, please refer to Figure S2 in the Supporting Information. Typically, FM films that are grown by sputtering maintain a polycrystalline structure. Indeed, several techniques can be adopted to induce the desired magnetic anisotropy in Co and Co-based thin films, such as the application of a magnetic field during the deposition, the tuning of the angle of deposition, or the introduction of external stress.[31–33] However, the bare sputtered Co films have been shown to possess a certain degree of anisotropy in various studies, also depending on the specific substrate used, a crucial aspect to consider.[25,34–37] According to the FMR theory, whereas only the IP uniaxial anisotropy is present and $\vec{H}_{ext}$ is applied along the easy or the hard axis, we can consider $H_{k1}(\varphi) = -H_{k2}(\varphi) = |H_k| = H_k$ in Eq. 2. In Eq. (2), the $M_{eff}$ term corresponds to the effective magnetization, that can be written as $4\pi M_{eff} = 4\pi M_s - \frac{2k_{2\perp}}{M_s} - \frac{2k_{2\parallel}}{M_s}$, where $M_s$ is the saturation magnetization, $K_{2\perp}$ the perpendicular magnetic anisotropy constant and $k_{2\parallel}$ the IP anisotropy constant. For a detailed calculation that justifies the previous assumptions, please refer to the Supporting Information.[38,39]

In Figure 3 (a) the evolution of the $f_{res}(H_{res})$ curves acquired along the hard axis (i.e. minimum value of $H_k$) of the Co layer are reported and fitted with Eq.2 for all the samples listed in Table 1. Here, we assume a gyromagnetic ratio of $\gamma = 1.98 \cdot 10^7 Hz/Oe$, which corresponds to a g-factor value of about 2.25.[35,40] As reported in Fig. S3 of the Supplementary Information, the angular dependence of the $M_{eff}$ and $H_k$ parameters is acquired for all the studied samples. A negligible angular dependence of $M_{eff}$ is observed, with a range of variation between 0.5% and 6% for all the set, as modulated by the $k_{2\parallel}$ term. This is compatible with a constant and more relevant $k_{2\perp}$ value, as expected for a fixed FM thickness. Differently, a significant reduction of the $H_k$ value is observed as the thickness of MnPtSb increases. This indicates a stronger IP uniaxial anisotropy in the thicker films, which is consistent with a more organized Co crystalline structure. This scenario is fully compatible with the XRD and TEM results (see Figures 1 and 2), confirming that the partial Mn segregation in the thinner MnPtSb samples may affect the crystalline quality of the film, which is finally reflected on the top-Co layer's magnetic properties.

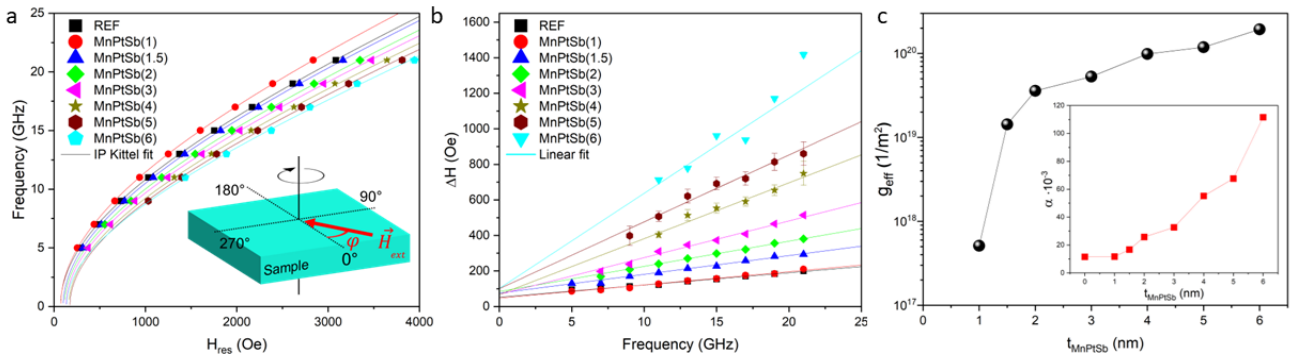

**Figure 3:** BFMR measurements performed by varying the direction of $\vec{H}_{ext}$ in the plane of the sample surface (see inset of panel (a)). In panel (a) the $f_{res}(H_{res})$ data fitted with Eq. 2., and in panel (b) the $\Delta H(f_{res})$ data fitted with Eq.3 are reported for all the samples summarized in Table 1. In panel (c) the $g_{eff}^{\uparrow\downarrow}$ as a function of the MnPtSb thickness is reported as extracted from Eq. 4. Here, the inset represents the evolution of the damping constant $\alpha$ for the different substrates' thicknesses.

Table 1: Summary of the prepared set of samples along with some of the main quantities of interest corresponding to each sample. The reported parameters are extracted along the hard axis of the Co magnetization vector.

| Label | Stack | $\Delta H_0$ (Oe) | $\alpha$ ($\cdot 10^{-3}$) | $g_{eff}^{\uparrow\downarrow}$ ($m^{-2}$) | $R_s$ ($\Omega$) | $W$ ($mm$) | $J_c^{2D}/h_{rf}$ ($mA \cdot m^{-1}$) |
|---|---|---|---|---|---|---|---|
| REF | Al$_2$O$_3$/Co(5nm)/Au(5nm) | 52 ± 2 | 11,51 ± 0,33 | / | 16 | 2.00 | (6,95 ± 0,1) $\cdot 10^{-5}$ |
| MnPtSb(1) | MnPtSb(1nm)/Co(5nm)/Au(5nm) | 47 ± 3 | 11,67 ± 0,38 | 5,12 $\cdot 10^{17}$ | 17 | 1.78 | (2,78 ± 0,03) $\cdot 10^{-4}$ |
| MnPtSb(1.5) | MnPtSb(1.5nm)/Co(5nm)/Au(5nm) | 75 ± 2 | 16,71 ± 0,44 | 1,43 $\cdot 10^{19}$ | 20 | 1.50 | (1,62 ± 0,02) $\cdot 10^{-4}$ |
| MnPtSb(2) | MnPtSb(2nm)/Co(5nm)/Au(5nm) | 84 ± 5 | 25,74 ± 0,84 | 3,58 $\cdot 10^{19}$ | 19 | 1.80 | (1,40 ± 0,01) $\cdot 10^{-4}$ |
| MnPtSb(3) | MnPtSb(3nm)/Co(5nm)/Au(5nm) | 68 ± 10 | 32,63 ± 0,88 | 5,28 $\cdot 10^{19}$ | 18 | 1.60 | (2,24 ± 0,01) $\cdot 10^{-4}$ |
| MnPtSb(4) | MnPtSb(4nm)/Co(5nm)/Au(5nm) | 67 ± 49 | 55,18 ± 4,45 | 9,85 $\cdot 10^{19}$ | 17 | 1.9 | (1,22 ± 0,01) $\cdot 10^{-4}$ |
| MnPtSb(5) | MnPtSb(5nm)/Co(5nm)/Au(5nm) | 100 ± 66 | 67,63 ± 7,09 | 1,18 $\cdot 10^{20}$ | 17 | 2.3 | (7,71 ± 0,13) $\cdot 10^{-5}$ |
| MnPtSb(6) | MnPtSb(6nm)/Co(5nm)/Au(5nm) | 102 ± 16 | 111,61 ± 9,22 | 1,94 $\cdot 10^{20}$ | 15 | 1.8 | (7,14 ± 0,14) $\cdot 10^{-5}$ |

Being the MnPtSb compound an FM material in itself, we also conducted BFMR on an additional Al$_2$O$_3$/MnPtSb(6)/Si(3) sample, i.e. without Co capping, in order to study its possible influence on the results obtained for the set of samples in Table 1. A comparison between the BFMR signals acquired for the Al$_2$O$_3$/MnPtSb(6)/Si(3) and MnPtSb(6) samples is reported in Fig. S4 of the Supplementary Information, where a very low intensity and a remarkably different resonant field are registered for the Co-free sample. Therefore, we conclude about the absence of any effect coming from the MnPtSb layers in the whole set of BFMR results summarized in Fig. 3.

In an IP BFMR measurement, the full width at half maximum ($\Delta H$) of a signal is expressed as a function of $f_{res}$ according to Eq. 3.

$$\Delta H = \Delta H_0 + \frac{4\pi}{|\gamma|} \alpha f_{res} \qquad [3]$$

where $\Delta H_0$ is the inhomogeneous broadening, accounting for the magneto-structural disorder of the FM film (i.e. structural disorder, magnetic dead layers). The evolution of $\Delta H_0$ and $\alpha$ as a function of $\varphi$ reflects the IP magnetic anisotropy of the Co film, despite Eq. 3 is not directly dependent by $\varphi$. Indeed, we plotted the angular evolution of the extracted $\Delta H_0$ and $\alpha$ values for each sample, demonstrating that for each MnPtSb thickness it is possible to identify two regimes characterized by high and low $\alpha$ values. In particular, the higher $\alpha$ value is obtained when $\vec{H}_{ext}$ is aligned along the hard axis, and the lower one when $\vec{H}_{ext}$ lies along the easy axis. For this reason, the effect of the MnPtSb layer on the magnetization dynamics in the Co layer was investigated by extracting the quantities of interest orienting the samples in the same IP crystalline direction. The exact positioning of the sample during the BFMR measurement can be obtained by performing IP XRD measurements and correlating the diffraction peaks with the extracted $\alpha$ values as a function of the IP angle (see Fig. S5 in the Supplementary Information).

In Fig. 3(b) the $\Delta H$ values extracted as a function of $f_{res}$ for all the samples, fixing the orientation of $\vec{H}_{ext}$ along the hard axis of the magnetization. Here, the solid lines indicate the fit of the dataset with Eq.3, from which we extracted the $\Delta H_0$ and $\alpha$ values for each sample (see Table 1). A progressive enhancement of the damping constant is observed with the increasing of the MnPtSb thickness, together with a similar trend for the $\Delta H_0$ values. The higher $\Delta H_0$ values suggest a higher degree of magneto-structural disorder in the Co film. However, the broader peaks measured for thicker samples makes the quantification of the FMR signal challenging at lower frequencies (below 11 GHz). This limitation introduces a significant source of uncertainty in the measurement, making it difficult to establish a straightforward correlation between the extracted $\Delta H_0$ values and the quality of the FM material in the thicker MnPtSb samples. Nevertheless, considering the raw numbers, the $\Delta H_0$ values lie within the 50-100 Oe range, which indicates an overall good magnetic quality of the samples.

According to the SP theory, the enhancement of $\alpha$ is directly related to the generation of pure spin currents in a FM layer.[41,42] During a BFMR experiment, when a spin sink material is in contact with a FM, the generated spin currents inside the FM layer are absorbed from the spin sink, producing an enhancement of the measured damping constant. Such an enhancement is related to the *spin mixing conductance* $g_{eff}^{\uparrow\downarrow}$, a complex quantity whose real part is proportional to the amount of the spin current density ($J_S^{3D}$) crossing the FM/spin-sink interface and absorbed by the spin sink layer, as regulated by Eqs. 4 and 5.

$$Re(g_{eff}^{\uparrow\downarrow}) = \frac{4\pi M_s t_{FM}}{g\mu_B}(\alpha_{S2} - \alpha_{S1}) \qquad [4]$$

$$J_S^{3D} = \frac{Re(g_{eff}^{\uparrow\downarrow})\gamma^2 h_{RF}^2 \hbar}{8\pi\alpha^2}\left(\frac{4\pi M_S\gamma - \sqrt{(4\pi M_S\gamma)^2 + 4\omega^2}}{(4\pi M_S\gamma)^2 + 4\omega^2}\right)\frac{2e}{\hbar} \qquad [5]$$

where $h_{RF}$ is the transverse magnetic field generated by the RF current, $t_{FM}$ the thickness of the FM layer and $\omega = f/2\pi$ is the angular RF frequency (see Methods). As it emerges from Eq. 4, $g_{eff}^{\uparrow\downarrow}$ is independent from the measuring condition (i.e., RF-frequency, RF-power), therefore representing an intrinsic property of the system. In view of the latter considerations, it is crucial to stress about the importance of choosing a proper reference sample (REF in Table 1), in order to correctly quantify the generated net spin current $J_S^{3D}$. Based on Eqs. 4 and 5 the higher is $\alpha$ the larger is the generated $J_S^{3D}$, justifying our choice to study the whole heterostructure when the Co layer is aligned along its hard axis, where the $\alpha$ value is maximized.

In Figure 3(c), we show the extracted $g_{eff}^{\uparrow\downarrow}$ values as a function of the MnPtSb thickness, ranging from about $1 \cdot 10^{21}\ m^{-2}$ for the thicker MnPtSb(6) down to $4 \cdot 10^{17} m^{-2}$ for the thinner MnPtSb(1). This variation of about four orders of magnitude in the $g_{eff}^{\uparrow\downarrow}$ values can be explained by the lower crystalline and morphological quality of the thinner MnPtSb layers, which are characterized by less sharp interfaces, a different degree of crystallization and stoichiometry as demonstrated by the XRD analysis (see Fig. 1), most likely preventing an optimized spin transport across the Co/MnPtSb interface. The $g_{eff}^{\uparrow\downarrow}$ values extracted from the studied samples are compatible with those calculated for many spintronic materials (Ref. [43]).

In SP-FMR measurements conducted as a function of the thickness of the spin-sink layer (i.e. MnPtSb), backflow effects of the generated spin currents must be considered in the case of SCC arising from bulk mechanisms, such as the Inverse Spin Hall Effect (ISHE).[44,45] In the latter case, the value of $\alpha$ progressively drops when the thickness is approaching the value of the spin diffusion length $\lambda_S$ for that specific material. This effect can be observed by plotting the $\alpha(t_{SS})$ curve, which will be characterized by a plateau at thicknesses higher than $\lambda_S$.[45–47] As shown in the inset of Fig. 3(c), in our samples we did not observe a similar behavior. Instead, we observe a parabolic type of dependence of $\alpha$ on $t_{MnPtSb}$. Even though $\alpha$ increases with the increase of the MnPtSb thickness, this finding suggests that our data are not in accordance with the existence of ISHE.

**Electrically detected spin pumping**

To gain more insight into the SCC conversion mechanisms occurring in our MnPtSb/Co-based systems, we measure the voltage drop arising at the edges of all the samples listed in Table 1. Such a signal arises due to the injection of pure spin currents from the Co layer into the MnPtSb one. Here, the same methodological approach used for the BFMR measurement is adopted (i.e. orientation of the sample with respect $\vec{H}_{ext}$) (see Methods).

Conducting electrically detected SP-FMR measurements allows to establish if the $J_S^{3D}$ generated in the Co layer is converted into a charge current due to the coupling with the MnPtSb layer, enabling the quantification of the SCC efficiency in such a system. Consequently, similar electrical measurements (i.e. spin torque FMR or II harmonic Hall signal[48]) are demanding for a thorough and reliable description of the spin-charge conversion phenomena.[43,49]

The DC voltage occurring at the edges of a sample due to SP is usually called *mixing voltage* ($V_{mix}$), being the convolution of different phenomena. In the framework of the SP theory, $V_{mix}$ can be fitted using a linear combination of Lorentzian functions as in the following Eq. 5.[50,51]

$$V_{mix} = V_{Sym} \frac{\Delta H^2}{\Delta H^2 + (H - H_{res})^2} + V_{Asym} \frac{\Delta H(H - H_{res})}{\Delta H^2 + (H - H_{res})^2} \quad (5)$$

where $V_{Sym}$ and $V_{Asym}$ represent the symmetric and anti-symmetric components of the global $V_{mix}$ signal, which are interpreted as the SP and the spin rectification effects (SRE) contributions (i.e. anisotropic magnetoresistance, anisotropic Hall effect), respectively. In Figure 4(a) the SP signals are displayed as obtained for MnPtSb(1) (black circles) for positive and negative magnetic fields maintaining the sample in the IP configuration, where the SP contribution is maximum. Here, the $V_{mix}$ (red solid line) signal is separated into the $V_{Sym}$ (green solid line) and $V_{Asym}$ (blue solid line) components through Equation 5, with the aim to isolate the SP effect. The higher intensity of $V_{Sym}$ with respect to $V_{Asym}$ indicates that the SRE are negligible. This suggests that the transport of spin angular momentum from the MnPtSb into the Co layer is the main cause of the electrical response of the system. Moreover, the reversal of the $V_{Sym}$ component with the inversion of the external magnetic field indicates the spin-dependent origin of such electrical signal, confirming the expectation. The symmetric component for positive and negative applied magnetic fields exhibits similar intensity, indicating the absence of thermal contributions (i.e. Seebeck effect).[52] In Figure 4 (b) the $V_{Sym}$ and $V_{Asym}$ components extracted from the MnPtSb(1) are reported as a function of the RF power, showing their linear evolution, a behavior that is compatible with the SP-FMR measurements.[44] The same properties described above for MnPtSb(1) can be found in all the samples reported in Table 1, thus constituting a reliable set for the study of SCC phenomena in the MnPtSb compound. In order to extract the SP voltage ($V_{SP}$) free from eventual thermal effects, the relation $V_{SP} = \frac{V_{Sym}(+H_{ext}) - V_{Asym}(-H_{ext})}{2}$ is used. In Figure 4 (c), the charge current density ($J_C^{2D}$) normalized over the strength of the oscillating magnetic field ($h_{RF}$) is reported for each sample, being $\frac{J_{SP}}{h_{RF}} = \frac{1}{h_{RF}} \frac{V_{SP}}{R_s W}$, where $R_s$ and W are the sheet resistance (i.e. four point measurements) and the sample width, respectively. A summary of the electrical parameters is reported in Table 1 for each studied heterostructure. From Fig. 4(c) it emerges a nearly descendent monotonic behavior of the $\frac{J_{SP}}{h_{RF}}$ curve as a function of the MnPtSb thickness, with the minimum value achieved for the MnPtSb(6) heterostructure being comparable with the MnPtSb free reference structure (i.e. REF). On the contrary, a significant increase of the generated electrical signal is observed for the thinnest MnPtSb(1)-based heterostructure. Such a behavior clearly points towards a SCC mechanism originating from the MnPtSb/Co interface, which progressively lowers when increasing the MnPtSb thickness. Such a surface behavior is in accordance with the BFMR results, where we do not observe plateau in the $\alpha(t_{MnPtSb})$ dependence, which must be present in the case of ISHE as the origin of the SCC, see inset in Fig. 3(c).[45–47]

As thoroughly discussed by Z. When et al.[17], in the fully spin polarized stoichiometric half-HC compounds, the generation of spin current should be prevented within the SP theory, because the creation of opposite spin's sign at different regions in these type of compounds is theoretically forbidden.[19,20,50] Therefore, a possible reason for explaining the observation of such a SCC in a half-HC, can be the deviation from its perfect stoichiometry taking place in the close vicinity of the MnPtSb/Co interface. This is exactly the scenario that we can draw according to the TEM and EDX results presented in Section 1b, where a partial Mn-depletion occurring in MnPtSb in the region closer to the Co layers occurs (see Fig. 2). Therefore, the MnPtSb/Co interfacial region can become active in terms of a possible SCC in our half-HC, a process becoming more relevant in thinner films, where the Mn-depletion has a larger impact.

Additionally, we performed temperature dependent (down to 5 K) SP-FMR on the thinnest MnPtSb(1)-based heterostructure, which support the surface-like SCC (see Fig. S6 in the Supporting Information).[20]

Having identified the Co/MnPtSb interface, i.e. the MnPtSb surface, as the region where the SCC occurs, we can calculate the SCC efficiency within the IEE model, where the figure of merit is the so-called IEE length $\lambda_{IEE}$. For the MnPtSb(1) heterostructure, we find a $\lambda_{IEE}$ = 3 nm, at room temperature. This value is surprisingly very high as compared the common values found in literature, which often lie well below 1 nm.[42] To our knowledge, the highest $\lambda_{IEE}$ so far reported at room temperature is 2.1 nm, which is measured in the $\alpha$-Sn system studied in Ref.[53].

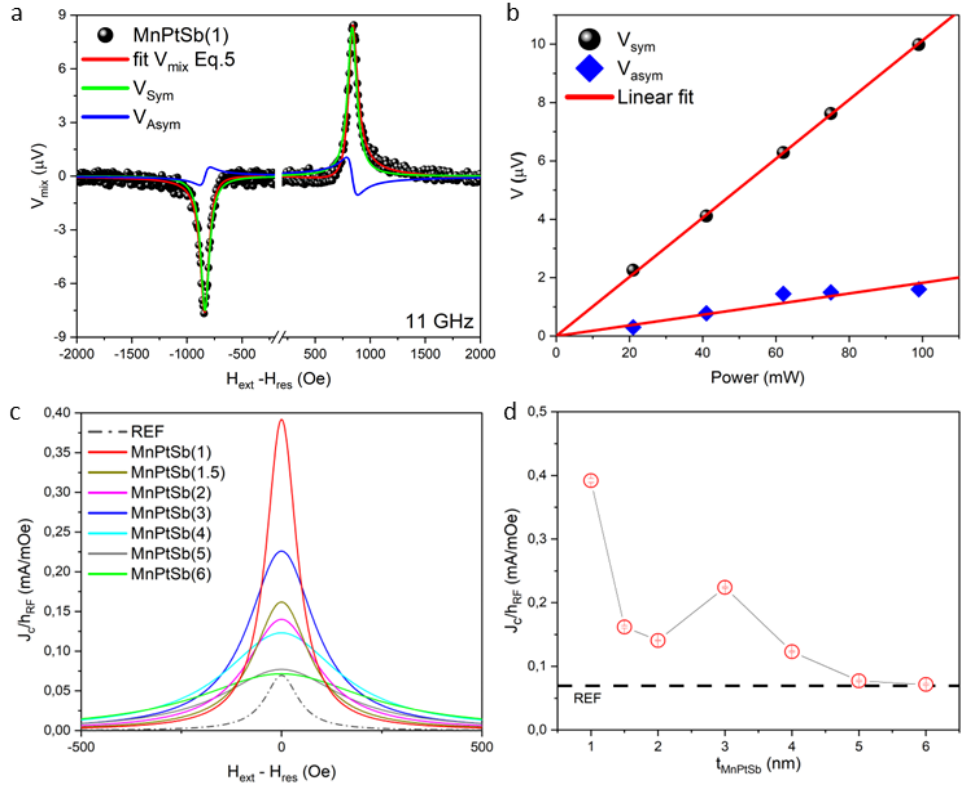

**Figure 4:** (a) V$_{mix}$ signal acquired for MnPtSb(1) at 11GHz and a microwave power of 99 mW ($h_{RF} = 0.79\ Oe$). The collected data are fitted with Eq. 5 (red solid line), from which the $V_{Sym}$ (green solid line) and $V_{Asym}$ (blue solid line) components are extracted. (b) $V_{Sym}$ and $V_{Asym}$ as a function of the RF power. The red solid lines show the linearity of the datasets, as expected in SP-FMR experiments. Panels (c) and (d) represents the charge current density normalized over the RF power extracted for all the samples reported in Table 1 as a function of $H_{ext} - H_{res}$ and the MnPtSb thickness, respectively.

**Conclusions**

In this manuscript, we report the epitaxial growth ultra-thin MnPtSb half Heusler compound and its spin-charge conversion (SCC) efficiency. The combination of X-ray diffraction and TEM measurements is employed to investigate the chemical-structural nature of the MnPtSb layers, which are nearly epitaxial in the thickest regime here reported (6 nm). In the case of thinner layers (< 2 nm), a partial Mn segregation is observed, which promotes the formation of a "PtSb"-rich phase in contact with Co. The magnetization dynamics and SCC mechanisms in MnPtSb/Co/Au heterostructures are investigated by FMR and SP-FMR measurements. We reveal the presence of in-plane uniaxial anisotropy in the Co films and a strong MnPtSb thickness dependent SCC effect. In particular, a remarkable enhancement of SCC is observed when reducing the MnPtSb thickness, demonstrating that the SCC is driven by the Co/MnPtSb interface. We attribute such an enhancement to the Mn-depletion observed by TEM in the thinner MnPtSb layers, which moves the MnPtSb away from its ideal half-metallicity, thus allowing the otherwise forbidden SCC. Within the inverse Edelstein effect model, we extract the record value of $\lambda_{IEE}$ = 3 nm at room temperature. Our results represent an important step towards the use of half-Heusler compounds for SCC applications, being the first direct quantification of their SCC efficiency for the MnPtSb compound.

**Materials and Methods**

MnPtSb (1 to 6 nm/Co (5nm)/Au (5nm) heterostructures were grown on (0011)-oriented $Al_2O_3$ single-crystal substrate using a BESTEC ultra-high vacuum magnetron sputtering system. Before deposition, the chamber was evacuated to a base pressure of less than $5 \times 10^{-8}$ mbar, whereas the process gas (Ar 5 N) pressure was $3 \times 10^{-3}$ mbar. The target-to-substrate distance was 20 cm, and the substrate was rotated at 20 rpm to ensure homogeneous growth. For the growth of the MnPtSb compound, we used Mn (5.08 cm), Pt (5.08 cm) and Sb (5.08 cm) sources in confocal geometry by applying 23 W, 15 W and 10 W DC power, respectively. The films were grown at 550 °C followed by an additional 30 min in-situ post annealing at the same temperature to improve the crystallinity. After the MnPtSb growth, we have grown Co and Au layers at room temperature. We used Co (5.08 cm) and Au (5.08 cm) sources in confocal geometry by applying 50 W and 20 W DC power, respectively. The growth rates and the film thicknesses were determined by using x-ray reflectivity (XRR) measurements.

The X-ray Diffraction (XRD) studies performed in the Bragg-Brentano geometry are performed with a HRXRD IS2000 diffractometer equipped with a Cu $K_\alpha$ radiation source (λ = 1.5406 Å), a four-circle goniometer, and a curved 120° position-sensitive detector (Inel CPS-120). This configuration allows to quantify the mosaicity of the film crystalline planes extracting information from the asymmetric reflections of the not perfectly parallel to the surface detection of the asymmetric reflections produced by the crystalline planes not perfectly parallel to the sample surface, giving access to the value of the mosaicity of the crystalline grains composing the material.

TEM measurements on electron transparent thin lamellae were performed with a Thermo Fisher Themis Z aberration-corrected Scanning Transmission Electron Microscope (STEM), operating at 200 kV acceleration voltage, equipped with an electron gun monochromator and a Dual-X Thermo Fisher energy dispersive X-rays (EDX) detector for EDX characterization. Morphological analyses were conducted with Bright Field TEM (BF-TEM) micrographs, while EDX elemental maps and profile analyses were acquired for high-resolution chemical characterization. To limit the electron beam damage, all TEM/STEM images and EDX maps were acquired with a low beam current. All the lamellae for TEM characterization were prepared using a Thermo Scientific™ Helios™ 5 UC DualBeam, which combines an Elstar electron column with a Tomahawk HT Focused Ion Beam (FIB). This preparation was made through in-situ extraction method with a tungsten micromanipulator system. To avoid the silicon amorphization final cleaning at 5 kV was applied.

BFMR is performed using a broadband Anritsu-MG3694C power source (1-40 GHz), connected to a grounded coplanar waveguide, where the samples are mounted in a flip-chip configuration (the FM film is located close to the GCPW surface), with a 75 µm thick Kapton foil stacked in between to prevent the shortening of the conduction line. The sample-GCPW system is positioned between the polar extensions of a Bruker ER-200 electromagnet maintaining its surface parallel to the external magnetic field $H_{ext}$, in the so-called in-plane (IP) configuration. During the measurements, an RF current at a fixed frequency is carried toward the GCPW and the transmitted signal is directed to a rectifying diode, converting the RF-signal in a continuous DC-current, subsequently detected by a lock-in amplifier downwards the electronic line. The same instrumentation is adopted to conduct SP-FMR measurements. Here, the edges of the sample are contacted with Ag paint and connected to a nanovoltmeter. A DC-voltage is detected in resonant condition, fixing the RF frequency and power. The temperature dependent measurements are performed with a Varian electromagnet equipped with a single frequency resonant cavity working in the X-band. In this case the sample was placed onto a quartz stick and electrically connected with Ag paint to the same nanovoltmeter used for the broadband measurements. A liquid He flow cryostat was exploited to cool down the sample.

**Acknowledgments**

We thank Edouard Lesne for discussions. We acknowledge the Horizon 2020 project SKYTOP "Skyrmion-Topological Insulator and Weyl Semimetal Technology" (FETPROACT-2018-01, n. 824123).

**Authors contribution**

E.L., A.M. and R.M. conceived the experiment. E.L. conducted all the *BFMR*, *SP-FMR* and *XRD* measurements. E.L. and M.B. developed the BFMR and SP-FMR set-up and performed the analysis of the FMR-based results. M.F. supervised the FMR activity at the University of Milano-Bicocca laboratory. A.M. produced the samples and performed magnetometry measurements. A.S., P.T. and D.C. performed and analyzed the TEM

# Supplementary Information

1. **Transmission Electron Microscopy (TEM) images of dedicated Al$_2$O$_3$/MnPtSb(6)/SiO$_x$(3), Al$_2$O$_3$/MnPtSb(2)/SiO$_x$(3)**

Dedicated MnPtSb films with nominal thicknesses of 6 nm and 3 nm capped with a 3 nm thick SiO$_x$ layer are grown for a first step TEM analysis. In Figure S1 the TEM images are shown for the two samples. Despite the films are electrically continuous, the TEM lamella evidences some island deposition, which makes difficult a precise quantification of the thickness of each layer in the full stack reported in Fig.2 of the main text. The MnPtSb islands can be better appreciated in the 6 nm thicker film and are labelled with red numbers in Fig. S1 (a). As it is clear from the island profile, they partially hide each other due to the tridimensional nature of the TEM lamella, from which islands belonging to regions far from the lamella surface appear close to each other. The Fast Fourier Transform (FFT) analysis of the crystalline structure shows the high quality of the films, as evident from the ordered pattern reported in the inset of Figures S1 (a) and (b). By measuring the thickness of the MnPtSb layers, a small deviation from the nominal layer thickness is observed, being slightly higher than the target values. However, in both samples the extra thickness is around the 25% of the nominal values (nominal thickness/extra thickness), a condition that can be considered preserved in all the samples studied in the main text, thus making the studied set of samples self-consistent.

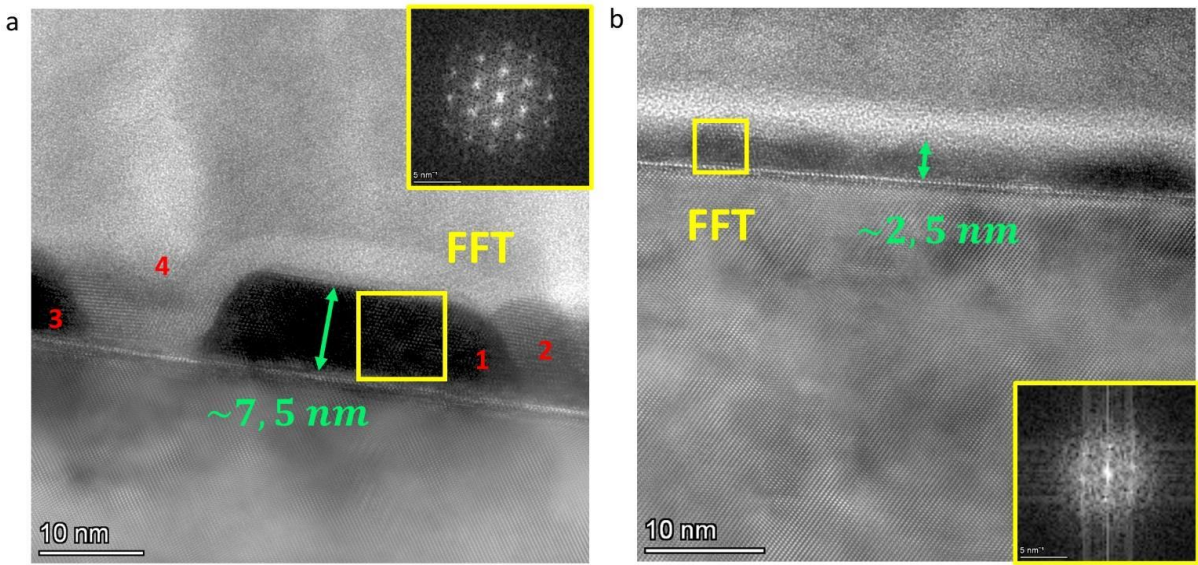

Figure S1: TEM images of the Al$_2$O$_3$/MnPtSb(6)/SiO$_x$(3) and Al$_2$O$_3$/MnPtSb(2)/SiO$_x$(3) heterostructures reported in panels (a) and (b), respectively. The island nature and the high crystallinity of the films is substantiated by qualitative considerations and FFT analysis.

2. **Evolution of the $H_{res}(\varphi)$ curves for the MnPtSb(1)/Co(5)/Au(5) sample**

As a first step to investigate the IP magnetic anisotropy present in the studied samples, the FMR signal was recorded in a resonating cavity for the fixed RF frequency of 9.4 GHz as a function of the IP angle for the MPS(1) sample. As emerges from Fig. S1, a 180° periodic oscillation is present by varying the IP angle, behavior compatible with the presence of a uniaxial magnetic anisotropy in the Co film. The collected data are fitted with the following equation

$$H_{res} = H_0 - H_k \cos(2(\varphi - \varphi_0))$$

The same angular dependence is observed in all the studied samples.

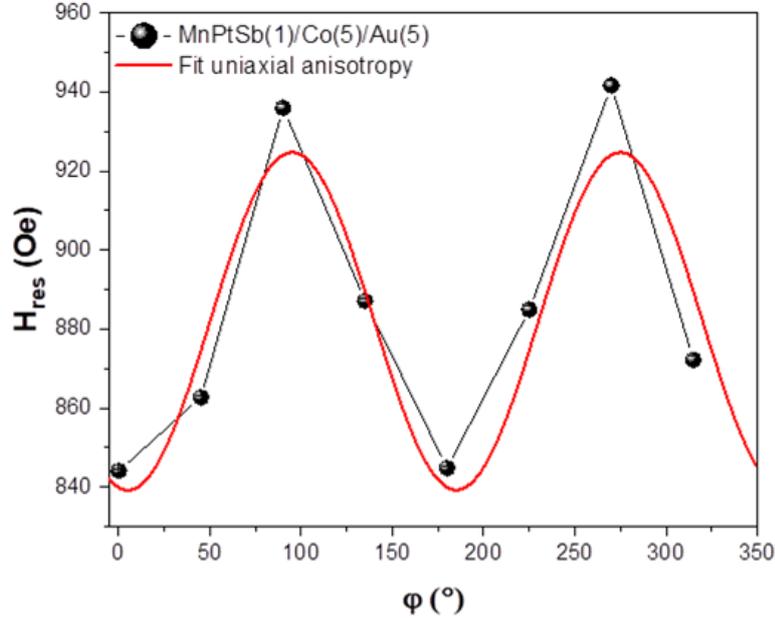

Figure S2: values as a function of the IP angle. The periodicity of the signal and the good quality of the fit demonstrate the presence of a uniaxial IP anisotropy in the deposited Co layers.

3. **Extraction the Kittel formula for the IP uniaxial anisotropy when the sample is aligned along the hard axis of the magnetization vector**

From the Smit-Beljiers equation we can write the resonant condition as

$$\left(\frac{\omega}{\gamma}\right)^2 = \frac{1}{(M_s \sin\theta_M)^2}\left(F_{\theta_M\theta_M}F_{\varphi_M\varphi_M} - F_{\theta_M\varphi_M}^2\right)$$

The free energy F of the magnetic system for the case of uniaxial in-plane (IP) and out-of-plane (OOP) magnetic anisotropy is

$$F = -\boldsymbol{M}\cdot\boldsymbol{H} + \frac{1}{2}\mu_0(N_\perp - N_\parallel)M^2\theta_M + F_{uniaxial}^{\parallel} + F_{uniaxial}^{\perp}$$

where

$$F_{uniaxial}^{\perp} = k_0 - k_{2\perp}\cos^2\theta_M \quad \text{(uniaxial OOP)}$$

$$F_{uniaxial}^{\parallel} = k_{2\parallel}\sin^2\theta_M \cos^2(\varphi_M - \delta) \quad \text{(uniaxial IP, } \delta \text{ defines the easy axis position)}$$

So, the full expression for F can be written as

$$F = -MH[\sin(\theta_H)\sin(\theta_M)\cos(\varphi_H - \varphi_M) + \cos(\theta_H)\cos(\theta_M)] + \frac{1}{2}\mu_0(N_\perp - N_\parallel)M^2\cos^2\theta_M$$
$$+ k_{2\parallel}\sin^2\theta_M \cos^2(\varphi_M - \delta) - k_{2\perp}\cos^2\theta_M$$

Calculating the following derivatives

$$\frac{dF}{d\theta_M} = -MH[\sin\theta_H \cos\theta_M \cos(\varphi_H - \varphi_M) - \sin\theta_M \cos\theta_H] - \frac{\mu_0}{2}(N_\perp - N_\parallel)M^2 \sin 2\theta_M$$
$$+ k_{2\parallel}\sin 2\theta_M \cos^2(\varphi_M - \delta) + k_{2\perp}\sin 2\theta_M$$

$$\frac{dF}{d\varphi_M} = -MH[\sin\theta_H \sin\theta_M \sin(\varphi_H - \varphi_M)] - k_{2\parallel} \sin^2\theta_M \sin(2\varphi_M - 2\delta)$$

$$\frac{d^2F}{d\theta_M^2} = +MH[\sin\theta_H \sin\theta_M \cos(\varphi_H - \varphi_M) + \cos\theta_M \cos\theta_H] - \frac{\mu_0}{2}(N_\perp - N_\parallel)M^2 2\cos 2\theta_M$$
$$+ k_{2\parallel} 2\cos 2\theta_M \cos^2(\varphi_M - \delta) + k_{2\perp} 2\cos 2\theta_M$$

$$\frac{d^2F}{d\varphi_M^2} = +MH[\sin\theta_H \sin\theta_M \cos(\varphi_H - \varphi_M)] - 2k_{2\parallel} \sin^2\theta_M \cos(2\varphi_M - 2\delta)$$

$$\frac{d^2F}{d\theta_M d\varphi} = -MH[\sin\theta_H \cos\theta_M \sin(\varphi_H - \varphi_M)] - k_{2\parallel} \sin 2\theta_M \sin(2\varphi_M - 2\delta)$$

Considering $\frac{dF}{d\theta_M}|_{eq} = 0$, $\frac{dF}{d\varphi_M}|_{eq} = 0$

and substituting in the derivative the condition $\theta_H = \theta_M = \frac{\pi}{2}$, we have

$$\begin{cases} \theta_M = \frac{\pi}{2} \\ -MH\sin(\varphi_H - \varphi_M) = k_{2\parallel} \sin(2\varphi_M - 2\delta) \end{cases}$$

Consequently, the Kittel equation can be written as

$$\left(\frac{\omega}{\gamma}\right)^2 = \left(H^{RES}\cos(\varphi_H - \varphi_M) + \mu_0(N_\perp - N_\parallel)M - \frac{2k_{2\perp}}{M} - \frac{2k_{2\parallel}}{M}\cos^2(\varphi_M - \delta)\right)\left(H^{RES}\cos(\varphi_H - \varphi_M)\right.$$
$$\left. - \frac{2k_{2\parallel}}{M}\cos(2\varphi_M - 2\delta)\right)$$

By adding and subtracting the term $\frac{2k_{2\parallel}}{M}\sin^2(\varphi_M - \delta)$ we can write

$$\left(\frac{\omega}{\gamma}\right)^2 = \left(H^{RES}\cos(\varphi_H - \varphi_M) + \mu_0(N_\perp - N_\parallel)M - \frac{2k_{2\perp}}{M} - \frac{2k_{2\parallel}}{M}\sin^2(\varphi_M - \delta) - \frac{2k_{2\parallel}}{M}\cos^2(\varphi_M - \delta)\right.$$
$$\left. + \frac{2k_{2\parallel}}{M}\sin^2(\varphi_M - \delta)\right)\left(H^{RES}\cos(\varphi_H - \varphi_M) - \frac{2k_{2\parallel}}{M}\cos^2(\varphi_M - \delta) + \frac{2k_{2\parallel}}{M}\sin^2(\varphi_M - \delta)\right)$$

Defining

$$\pi M_{eff}(\varphi_M) = +\mu_0(N_\perp - N_\parallel)M - \frac{2k_{2\perp}}{M} - \frac{2k_{2\parallel}}{M}\sin^2(\varphi_M - \delta)$$

$$H_k(\varphi_M) = -\frac{2k_{2\parallel}}{M}\cos^2(\varphi_M - \delta) + \frac{2k_{2\parallel}}{M}\sin^2(\varphi_M - \delta)$$

it means that at the easy axis we have

$$4\pi M_{eff} = +\mu_0(N_\perp - N_\parallel)M - \frac{2k_{2\perp}}{M}$$

$$H_k(\varphi_M) = -\frac{2k_{2\parallel}}{M}$$

and at the hard axis

$$4\pi M_{eff} = +\mu_0(N_\perp - N_\parallel)M - \frac{2k_{2\perp}}{M} - \frac{2k_{2\parallel}}{M}$$

$$H_k(\varphi_M) = +\frac{2k_{2\parallel}}{M}$$

According to the different definition of $4\pi M_{eff}$ and considering the opposite sign of $H_k(\varphi_M)$, when the sample is aligned along the easy or the hard axis the Kittel equation can be written in the form

$$\left(\frac{\omega}{\gamma}\right)^2 = \left(H^{RES}\cos(\varphi_H - \varphi_M) + 4\pi M_{eff}(\varphi_M) + H_k(\varphi_M)\right)\left(H^{RES}\cos(\varphi_H - \varphi_M) + H_k(\varphi_M)\right)$$

Which considering $\cos(\varphi_H - \varphi_M) = 1$ at the equilibrium turns out to be of the same shape of the Eq.2 adopted in the main text

$$\left(\frac{\omega}{\gamma}\right)^2 = \left(H^{RES} + 4\pi M_{eff} + H_k(\varphi_M)\right)\left(H^{RES} + H_k(\varphi_M)\right)$$

## 4. $M_{eff}$ and $H_k$ parameters extracted from BFRM measurements

The IP angular dependence of the $M_{eff}$ and $H_k$ parameters extracted from the fit of the curves reported in Fig. 4(a) of the main text is reported. As expected, a negligible modulation of the $M_{eff}$ value is observed, due to the presence of the $-\frac{2k_{2\parallel}}{M}$ term (see previous section here in the Supp. Info.). On the other hand, $H_k$ significantly varies with φ, because of the presence of two preferential orientations in the film. The direction characterized by a lower value of $H_k$ corresponds to the hard axis. On the contrary, the higher $H_k$ values indicate the easy axis of the magnetization vector.

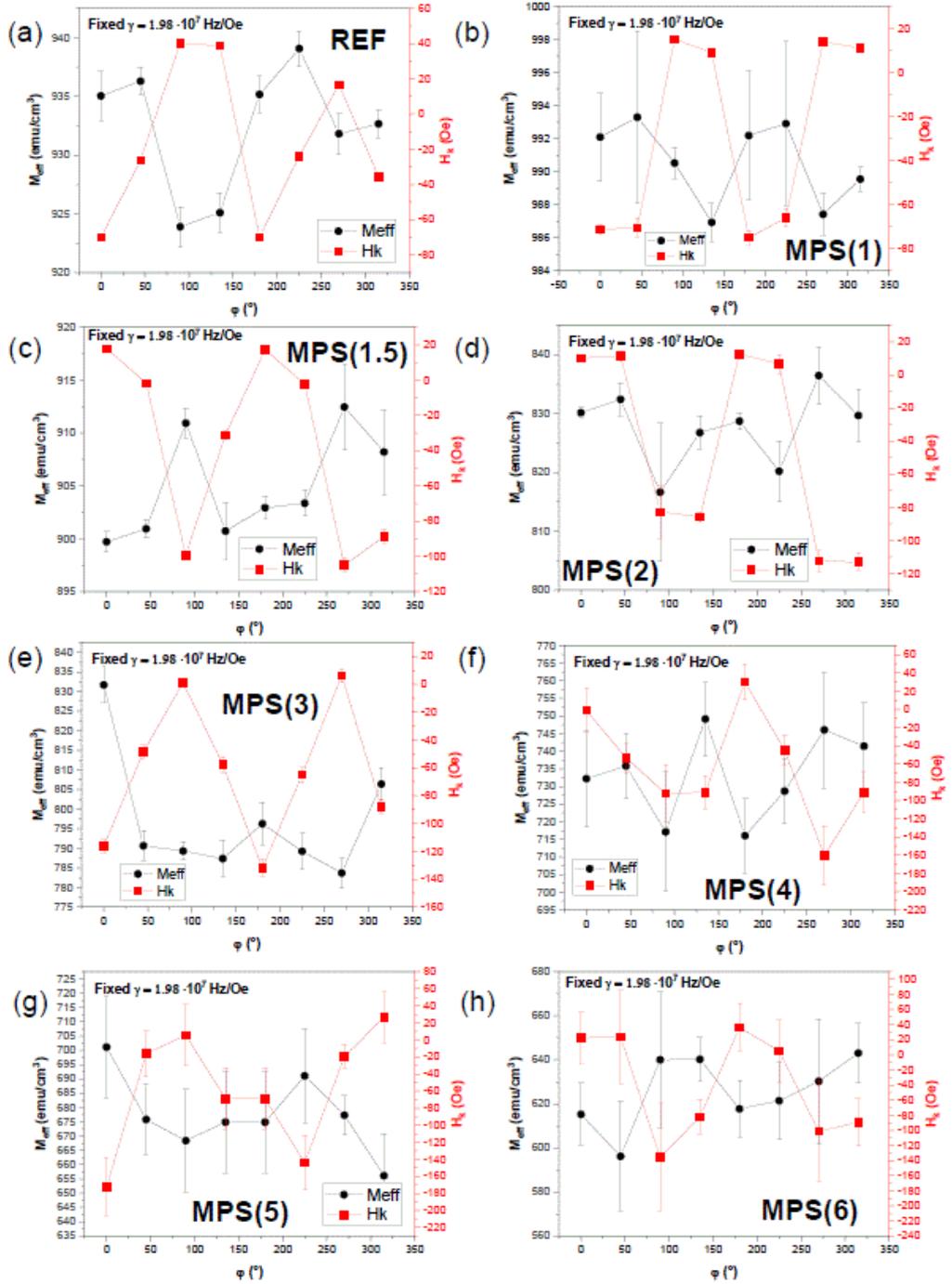

Figure S3: Evolution of the $M_{eff}$ and $H_k$ parameters extracted by the Kittel curves reported in Fig.4 of the main text.

## 5. FMR in-plane angular dependent measurements

The studied MnPtSb substrates exhibit a ferromagnetic nature. In order to disentangle the FMR signal of the MnPtSb layer from the one belonging to the Co layer, the same MnPtSb(6 nm)/Si(3 nm) sample shown in Fig. S1 (a) was compared with sample MPS(6) (see Table 1 in the main text). In Figure S4 a comparison between the FMR signals extracted from the latter samples is reported.

In Figure S4 (a) the FMR signal of the MnPtSb(6 nm)/Si(3 nm) sample is plotted by varying the orientation of the external magnetic field in the film plane. As it emerges from the dataset, no IP anisotropy is observed in the bare MnPtSb substrate. The FMR signal acquired at 0° is fitted with the following Lorentzian function describing the derivative of the power absorption (**P**) with respect the external magnetic field (**$H_{ext}$**)

$$\frac{dP}{dH_{ext}} = A \frac{4 \Delta H (H - H_{res})}{[4(H - H_{res})^2 + (\Delta H)^2]^2} - B \frac{(\Delta H)^2 - 4(H - H_{res})^2}{[4(H - H_{res})^2 + (\Delta H)^2]^2}$$

where $\Delta H$ and $H_{res}$ represent the signal linewidth and the resonant field, respectively. From the fit the values $\Delta H = 669 \, Oe$ and $H_{res} = 2231 \, Oe$ are extracted. The same study was performed for the FMR signal of sample MPS(6), obtaining $\Delta H = 331 \, Oe$ and $H_{res} = 917 \, Oe$.

In summary, for the same frequency the two FMR signals for the samples with and without Co are clearly distinguishable, having remarkably different linewidth and resonant field. The FMR signal acquired in Figure S4 (b) also has an opposite phase with respect to the signal in panel (a). Moreover, it must be noted that the signal reported in panel (a) is significantly weaker that the one in panel (b), indeed the former was acquired using a FMR resonating cavity, which is orders of magnitude more sensitive than the grounded coplanar waveguide adopted for the FMR studies discussed in the main text. The FMR signal of the MnPtSb(6 nm)/Si(3 nm) sample is not visible when GCPW is used, ensuring that we only followed the FMR signals originated from the Co layer.

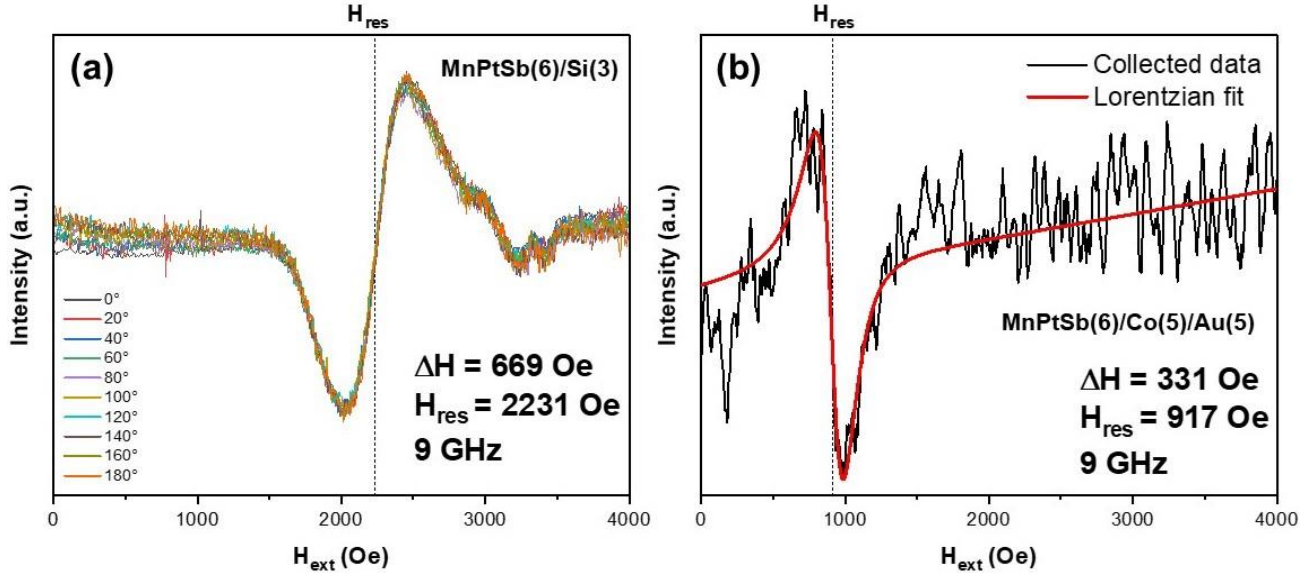

Figure S4: Comparison of the FMR signals arising from the MnPtSb(6 nm)/Si(3 nm) and MPS(6) sample for the same RF frequency.

## 6. *Correlation between the sapphire crystallographic orientation and the FMR measurements*

In Figure S5 (a) and (b) the in-plane XRD scan performed on samples MPS(6) and MPS(1) are reported, respectively. The aim of the measurement is to relate the oscillatory behavior of the broadband FMR response with respect to the orientation of the $Al_2O_3$ crystal. From the producer's datasheet, the $Al_2O_3$ substrates are C-plane oriented and cut in 1x1 cm² pieces, thus with the out-of-plane orientation along the [0001] direction and the edges along the [11-20] and [1-100] ones. To identify the orientation of the substrates, we performed in-plane XRD measurements aligning the samples along the asymmetric (104) crystalline plane ($2\vartheta \sim 35.2°$) and positioning one of their edges parallel to the X-ray beam, corresponding to the 0° position on the y-axes in Figures S5(a) and (b). As a result, two clear peaks spaced 120° emerged in Fig. S5 (a) for the MPS(6) sample at about $2\vartheta = 35.2°$, in accordance with the 3-fold symmetry of the (104) reflection. The same measurement performed for sample MPS(1) and reported in Fig. S5 (b) showed a shift of the (104) peak of about 30° with respect to the initial position (i.e. position of the sample edge parallel to the X-ray beam at 0°), indicating a rotation of 90° of the substrate with respect Fig. S5 (a) (the dashed white line indicates the 90° rotation).

To correlate the crystalline orientation with the angular evolution of the FMR signals, we performed FMR measurements on the two samples fixing the 0° position of the XRD measurement with the 0° FMR scan. What emerged is a correlation between the two measurements: the slope of the $\Delta H(f_{res})$ curve is lower if the FMR signal is measured along the (104) and higher when it is rotated by 90°. This determines an oscillatory behavior of the Slope(°) curve, identifying a high slope regime and a low slope regime, which according to Eq. 3 in the main text corresponds to high and low damping regimes. Such a behavior is related to the anisotropic nature of the crystalline structure of the Co layer deposited on top of the MnPtSb, which developed a (111) out-of-plane texturization during the growth. To extract the spin-charge conversion efficiency for the studied systems, in the main text we analyzed the results obtained for the high slope regime.

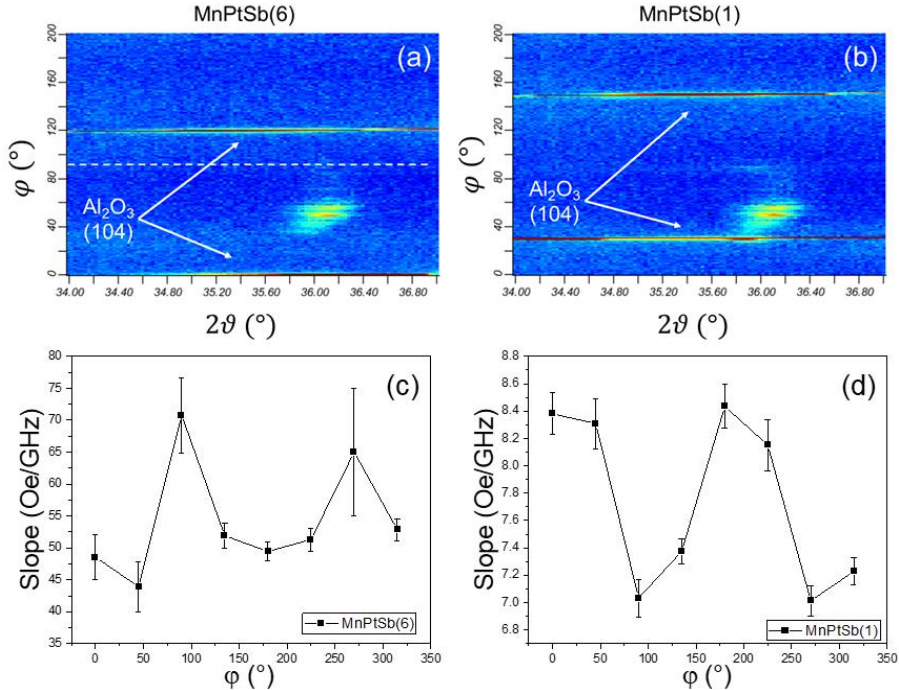

Figure S5: In panels (a) and (b) the in-plane XRD measurement centered around the (104) Al₂O₃ single crystal reflection are shown for samples MPS(6) and MPS(1), respectively. The white dashed line is a guide for the eyes indicating the 90° rotation around the out-of-plane axis between the two samples. In Panels (c) and (d) the evolution of the slope of the $\Delta H(f_{res})$ curve is reported as a function of the angle between the edge of the sample and the external applied magnetic field. The 0° on the x-axis corresponds to the 0° of the y-axes of panels (a) and (b).

### 7. Temperature dependent SP-FMR measurements

Temperature dependent SP-FMR measurements performed on the MPS(1) heterostructure (see Table 1 main text). These measurements are performed with a single frequency resonant cavity working in the X-band, using a standard cryostat with liquid He. In Fig. S6 (a) the $V_{mix}$ signal acquired for sample MPS(1) is reported as a function of the temperature down to 4 K. In Fig. S6 (b) the $V_{sym}$ component of each curve reported in panel (a) as extracted from Eq. 5 in the main text is plotted as a function of the temperature. According to the theory discussed in Ref. [SR1], the negative $V_{sym}$ value at T = 0 K indicates the presence of surface conductive states in the MnPtSb layer, thus supporting the interpretation of the data discussed in the main text and justifying our choice to adopt the Inverse Edelstein Effect model for the extraction of the spin-to-charge conversion efficiency in our systems.

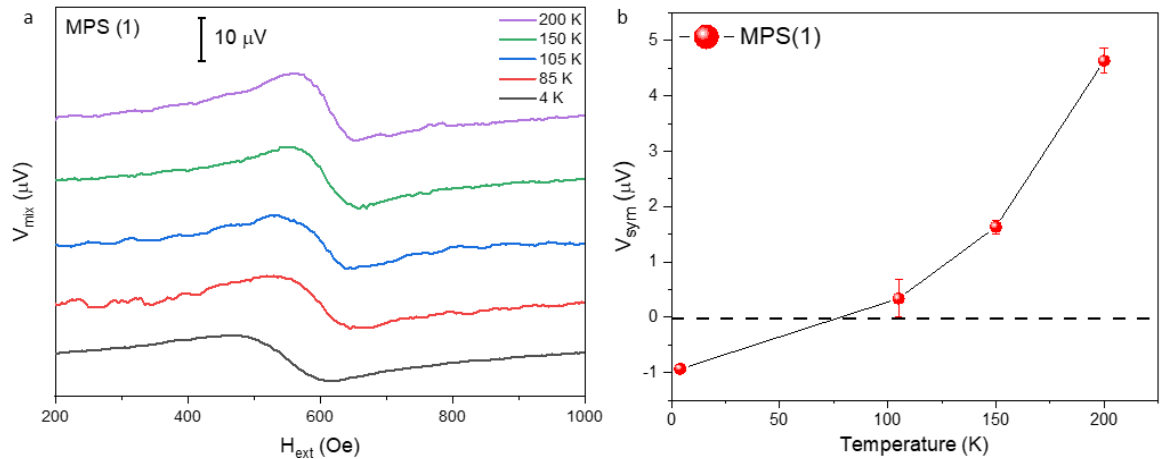

Figure S6: spin pumping signal as a function of the temperature for sample MPS(1). (b) intensity of the symmetric component of each curve reported in panel (a) as a function of the temperature for the same sample. The black solid line is a guide for the eyes.